\def\year{2021}\relax
\documentclass[letterpaper]{article} 
\usepackage{aaai21}  
\usepackage{times}  
\usepackage{helvet} 
\usepackage{courier}  
\usepackage[hyphens]{url}  
\usepackage{graphicx} 
\usepackage{natbib}  
\usepackage{caption} 
\usepackage{url}                                
\usepackage{array,multirow,graphicx,adjustbox}  
\usepackage{booktabs}                           
\usepackage{color}
\usepackage{subfigure}
\usepackage{paralist}
\usepackage{enumitem}

\newcommand{\descr}[1]{\smallskip\noindent\textbf{#1}}

\renewenvironment{thebibliography}[1]{
 \begin{oldthebibliography}{#1}
   \setlength{\itemsep}{0.1em}
   \setlength{\parskip}{0.1em}
}
{
 \end{oldthebibliography}
}

\renewcommand{\footnoterule}{%
 \kern -3pt
 \hrule width 1in 
 \kern 2pt
}

\nocopyright

\graphicspath{{./figs/}}

\title{On the Globalization of the QAnon Conspiracy Theory Through Telegram}
\author {
    Mohamad Hoseini,\textsuperscript{\rm 1,\thanks{Authors contributed equally}}
    Philipe Melo, \textsuperscript{\rm 2, \footnotemark[1]}
    Fabricio Benevenuto, \textsuperscript{\rm 2}\\
    Anja Feldmann, \textsuperscript{\rm 1} 
    Savvas Zannettou \textsuperscript{\rm{1}}\\
}
\affiliations {
    \textsuperscript{\rm 1} Max Planck Institute for Informatics \\
    \textsuperscript{\rm 2} Federal University of Minas Gerais \\
    \{mhoseini, anja, szannett\}@mpi-inf.mpg.de, \{philmelo, fabricio\}@dcc.ufmg.br
}

\begin{document}
\maketitle

\begin{abstract}

QAnon is a far-right conspiracy theory that became popular and mainstream over the past few years.
Worryingly, the QAnon conspiracy theory has implications in the real world, with supporters of the theory participating in real-world violent acts like the US capitol attack in 2021.
At the same time, the QAnon theory started evolving into a global phenomenon by attracting followers across the globe and, in particular, in Europe.
Therefore, it is imperative to understand how the QAnon theory became a worldwide phenomenon and how this dissemination has been happening in the online space.

This paper performs a large-scale data analysis of QAnon through Telegram by collecting 4.5M messages posted in 161 QAnon groups/channels.
Using Google's Perspective API, we analyze the toxicity of QAnon content across languages and over time.
Also, using a BERT-based topic modeling approach, we analyze the QAnon discourse across multiple languages.
Among other things, we find that the German language is prevalent in QAnon groups/channels on Telegram, even overshadowing English after 2020.
Also, we find that content posted in German and Portuguese tends to be more toxic compared to English.
Our topic modeling indicates that QAnon supporters discuss various topics of interest within far-right movements, including world politics, conspiracy theories, COVID-19, and the anti-vaccination movement.
Taken all together, we perform the first multilingual study on QAnon through Telegram and paint a nuanced overview of the globalization of the QAnon theory.

\end{abstract}

\section{Introduction} \label{sec:intro}

The spread of conspiracy theories online is not a new problem.
For instance, conspiracy theories related to the 9/11 attack (e.g., the attack was, in fact, a controlled demolition) and the Sandy Hook shooting (e.g., the shooting was staged)  were extensively disseminated on the Web.
Over the past few years, however, we witness an explosion in the spread and popularity of conspiracy theories on the Web.
More worrying is that newer conspiracy theories seem to have a more negative impact on the online and offline world. 
People get radicalized online from the continuous exposure to conspiratorial content and then perpetrate violent acts in the real world.
For instance, the Pizzagate conspiracy theory was the driving factor for a shooting at a pizzeria in Washington DC in 2016~\cite{pizzagate_shooting}.
Taken all together, there is a pressing need to understand how these conspiracy theories spread online and how users are radicalized from the exposure to conspiratorial content.

One conspiracy theory that is considered very persuasive and attracted high engagement from people is the QAnon conspiracy theory.
The theory alleges that a secret cabal of people run a child sex-trafficking ring and is working against Donald Trump (as the US president by the time the theory emerged).
Over the years, the conspiracy theory attracted many new followers across the globe and essentially evolved into a cult. 
Worryingly, the followers of the QAnon conspiracy theory started making threats or participate in violent real-world incidents (e.g., Capitol attack in 2021~\cite{qanon_capitol}), hence highlighting the impact that the conspiracy theory has in the real world~\cite{qanon_violence}.

Motivated by the negative impact that QAnon has in the real world, mainstream platforms like Facebook, Twitter, and YouTube, started moderating and removing QAnon-related content.
Due to this, QAnon supporters sought new online ``homes'' in less-moderated platforms and migrated to other platforms like Parler and Telegram.  
At the same time, the QAnon conspiracy theory became a global phenomenon; the QAnon conspiracy theory started accumulating new followers across the globe, particularly in European countries like Germany and Spain~\cite{qanon_europe}.
Overall, it is imperative to understand how QAnon evolved and became a global phenomenon over time.
To do this, we use Telegram as the source of our study for two reasons. 
First, anecdotal evidence suggests that QAnon followers migrated to Telegram after bans on other platforms~\cite{telegram_qanon}.
Second, Telegram is a rapidly growing platform with worldwide coverage, hence it is the ideal platform for effectively studying the QAnon conspiracy theory across the globe.

\descr{Research Questions.} We focus on answering the following research questions:
\begin{itemize}[noitemsep,topsep=0pt,parsep=0pt,partopsep=0pt]
	\item \textbf{RQ1:} How does the QAnon community evolve on Telegram over time and across languages? How toxic is the QAnon community on Telegram?
	\item \textbf{RQ2:} How popular is QAnon content on Telegram?
	\item \textbf{RQ3:} What are the main topics of discussion on QAnon-related groups/channels on Telegram? Are there differences across languages?
\end{itemize}

To answer the above-mentioned research questions, we perform a large-scale data collection and analysis of QAnon-related groups/channels on Telegram.
Overall, we collect 4.5M messages shared in 161 Telegram groups/channels between September 2017 and March 2021.
Using Google's Perspective API, we investigate the toxicity of QAnon content on Telegram and assess whether the movement is becoming more toxic over time and whether there are substantial differences across languages.
Also, using a multilingual BERT-based topic modeling approach, we study the QAnon discourse across multiple countries/languages.

\descr{Main findings.} Our study provides some key findings:

\begin{itemize}[noitemsep,topsep=0pt,parsep=0pt,partopsep=0pt]
	\item We find that the QAnon movement increased substantially during 2020 and 2021 on Telegram, which is likely due to people migrating from other mainstream platforms that take moderation actions against QAnon content. Furthermore, by comparing content across languages, we find that German content overshadowed English in popularity during 2020 and 2021 (\textbf{RQ1}).
	\item By analyzing the toxicity of QAnon-related messages, we find that toxicity is rising over time (almost 2x times more toxic messages in March 2021 compared to September 2019), while at the same time, we find substantial differences across languages, with German and Portuguese being the most toxic languages. (\textbf{RQ1}).
	\item Our analysis indicates that QAnon content on Telegram is popular and can be viewed by many users. Specifically, we find one order of magnitude more views and forwards on QAnon content than a baseline dataset containing a set of political oriented groups. Also, we find that over time QAnon content is becoming even more popular and reaches an increasing number of Telegram users (\textbf{RQ2}).
	\item Our discourse analysis highlights that QAnon has evolved into discussing various topics of interest within far-right movements across the globe. We find several topics of discussions like world politics, conspiracy theories, COVID-19, and the anti-vaccination movement (\textbf{RQ3}).
\end{itemize}

\noindent Overall, our analysis portraits a nuanced overview of the QAnon movement across multiple countries.
Our observations suggest that the QAnon movement is adapting and expanding from the US to other countries, usually embodied in far-right movements.
As a consequence, efforts to reduce redundant work from fact checkers around the world, such as the \#CoronaVirusFacts led by the International Fact-Checking Network~\cite{covid_poynter}, may be relevant to debunk misinformation associated with QAnon globally.

\section{Background \& Related Work}
To the best of our knowledge, our effort is the first to explore the spread of QAnon movement over Telegram. Thus, next we briefly we provide background information and review previous work related to QAnon and Telegram. 

\descr{QAnon.} QAnon is a conspiracy theory alleging that a secret group of people (i.e., a cabal consisting of Democratic politicians, government officials, and Hollywood actors) were running a global child sex trafficking ring and were plotting against former US president Donald Trump~\cite{qanon_guardian}.
The conspiracy theory started in October 2017 with a post on 4chan by a user named ``Q,'' who claimed that he was an American government official with classified information about plots against then-President Donald Trump.
Subsequently, ``Q'' continued disseminating cryptic messages about the QAnon conspiracy theory (called ``Q drops'') mainly on 8chan. 
The QAnon conspiracy theory accumulated many followers on fringe Web communities like 4chan/8chan and mainstream ones like Facebook~\cite{qanon_fb_presence} and Twitter, especially after then-president Donald Trump retweeted QAnon-related content~\cite{qanon_twitter_trump}.
QAnon followers usually use their motto ``Where We Go One, We Go All'' (or simply wwg1wga) to tag content related to the QAnon conspiracy theory on the Web.

Over the past few years, followers of the QAnon conspiracy theory made violent threats or were linked with several incidents of real-world violence~\cite{qanon_violence}, with the Federal Bureau of Investigation (FBI) labeling it as a potential domestic terrorist threat~\cite{qanon_fbi_domestic}. 
The last straw of these incidents happened on January 6th, 2021, when supporters of the QAnon conspiracy theory attacked the US capitol in an attempt to overturn Donald Trump's defeat in the 2020 US elections by disrupting the Congress that was in the process of formalizing Joe Biden's victory~\cite{qanon_capitol}.
Due to these threats and violent incidents, mainstream platforms like Facebook~\cite{facebook_suspensions}, Twitter~\cite{twitter_suspensions}, Reddit~\cite{reddit_suspensions}, and YouTube~\cite{youtube_suspensions} started monitoring content related to QAnon and actively removing groups, subreddits, and users that are related to the QAnon conspiracy theory.
Naturally, following these content moderation interventions, supporters of the QAnon conspiracy theory flocked to other fringe Web communities, with lax moderation, like Parler~\cite{aliapoulios2021early} and Gab~\cite{lima2018inside,zannettou2018gab}, or messaging platforms like Telegram~\cite{telegram_qanon}.

Despite that the idea of the QAnon conspiracy theory is US-centric, the conspiracy theory recently started becoming a global phenomenon, and in particular it became popular among people with far-right ideology.
During 2020, the QAnon theory spread to Europe~\cite{qanon_europe}.
The conspiracy theory is nowadays shared among people from Spain, Italy, United Kingdom, and Germany, one of the most popular ``representatives'' in Europe~\cite{qanon_germany}.

Previous work investigates several aspects of the QAnon conspiracy theory.
\cite{papasavva2020qoincidence} analyze content toxicity and narratives in a Qanon community on Voat, finding that discussions in popular communities on Voat are more toxic than in QAnon communities.
\cite{aliapoulios2021early} provide a dataset of 183M Parler posts, and they highlight that QAnon is one of the dominant topics on Parler.
\cite{miller2021characterizing} investigates a sample of QAnon-related comments on YouTube, highlighting the international nature of the movement.
\cite{garry2021qanon} explore QAnon supporters’ behavior in spreading disinformation on Gab and Telegram, finding that the dissemination of disinformation is one of the main reasons for the dramatic growth of QAnon conspiracy.
\cite{hannah2021qanon} also investigate the reasons for the growth of QAnon, finding that sharing and discussing Q drops is one of the main reasons.
\cite{chandler2020we} investigates how QAnon followers are influenced by Q drops, finding that Q drops focus on the perceived allies or enemies of QAnon. 
\cite{planck2020we} compares the QAnon community’s rhetoric with a mainstream conservative community on Twitter, finding that tweets posted by QAnon supporters are more violent.
\cite{aliapoulios2021gospel} investigate a dataset of 4.9K canonical Q drops from six aggregation sites. 
They identify inconsistencies among the drops shared across aggregation sites and demonstrate that the drops have multiple authors.
Finally,~\cite{ferrara2020characterizing} investigate 240M election-related tweets 
finding that 13\% of users spreading political conspiracies (including QAnon) are bots.

\descr{Telegram.} Telegram is a popular messaging platform, with over 500M monthly active users~\cite{telegram_users}. 
Users can create public and private chat rooms called \emph{channels} or \emph{groups}.
Channels support few-to-many communication, where only the creator and few administrators can post messages, while groups support many-to-many communication (all members can post messages).
On Telegram, groups and channels can have a large number of members, with a limit of 200,000 members for groups and an unlimited number of members in channels, hence Telegram is useful for disseminating information to a large group of people.
Telegram users can share messages in groups/channels, with Telegram supporting text, images, videos, audios, stickers, etc.
Also, users can forward messages between groups/channels, with Telegram showing an indication in its user interface that the message is forwarded and the source channel/group.

Due to its privacy policy and encrypted nature (i.e., ``all data is stored heavily encrypted''), Telegram attracted the interest of dangerous organizations like terrorists~\cite{Telegram-Terrorists} and far-right groups~\cite{Telegram-White-Supremacists}.
Given this history and use of Telegram, in this work, we study the QAnon conspiracy theory through the lens of the Telegram platform. 
Also, we select Telegram as it is popular across the globe, hence assisting us in studying the globalization of the QAnon conspiracy theory.

\section{Dataset}\label{sec:dataset}

\begin{table}[]
\small
\centering
\begin{tabular}{lllrr}
\hline
\textbf{Dataset}                & \textbf{Source}     & \textbf{\#Groups} & \multicolumn{1}{l}{\textbf{\#Senders}} & \multicolumn{1}{l}{\textbf{\#Messages}} \\ \hline
\multirow{3}{*}{\textbf{QAnon}} & \textbf{Twitter/FB} & 77                & 92,475                                 & 3,564,381                               \\
                                & \textbf{Forwarded}  & 84                & 84                                     & 942,998                                 \\ \cline{2-5} 
                                & \textbf{Total}      & 161               & 92,559                                 & 4,507,379                               \\ \hline
\textbf{Baseline}               & \textbf{Twitter/FB} & 869               & 200,591                                & 9,358,946                               \\ \hline
\end{tabular}%
\caption{Overview of our Telegram dataset.}
\label{tab:dataset}
\end{table}

An inherent challenge that exists when studying phenomena through platforms like Telegram is to discover groups/channels related to the topic of interest.
To overcome this challenge and discover groups/channels related to QAnon, we follow and build on top of the methodology by~\cite{hoseini2020demystifying}.
Specifically, we: 
1)~search on Twitter and Facebook for URLs to Telegram groups/channels;
2)~collect metadata for each group/channel;
3)~select groups/channels based on QAnon-related keywords. 
4)~manually assess the relevance of the selected groups/channels;
5)~join and collect all messages from all the discovered QAnon groups/channels;
and 6)~expand our QAnon groups/channels based on forwarded messages shared in already discovered QAnon groups/channels and repeat Step 5. 
Below, we elaborate on each step.

\descr{1. Discovering groups/channels.} We use Twitter and Facebook to discover Telegram groups/channels. 
For Twitter, we use the Search and Streaming API to collect tweets that include Telegram URLs, following the methodology by~\cite{hoseini2020demystifying}, while for Facebook, we use the Crowdtangle API to obtain posts including Telegram URLs~\cite{crowdtangle}.
For both data sources, we perform queries with three URL patterns obtained from~\cite{hoseini2020demystifying}: \emph{t.me}, \emph{telegram.me}, and \emph{telegram.org}.
We collect Twitter and Facebook posts, including URLs from the above patterns between April 8, 2020, and October 10, 2020, ultimately collecting a set of 5,488,596 tweets and 14,004,394 Facebook posts that include a set of 922,289 unique Telegram URLs.
Note that the Crowdtangle API tracks and provides data only from publicly available Groups and Pages (i.e., does not include posts from user timelines).

\descr{2. Collecting group/channel metadata.} Having discovered a set of Telegram URLs, we then use Telegram's Web client and obtain basic group/channel metadata from the URLs. These include:
a) Name of the group/channel;
b) Description of the group/channel;
c) Number of members; and
d) the type of the Telegram URL (i.e., if it corresponds to a channel or a group).

\descr{3. Selecting QAnon groups/channels.} The next step is to narrow down the set of groups/channels to the ones that mention QAnon.
To do this, we search for the appearance of QAnon-related keywords on Twitter/Facebook posts that shared Telegram URLs or on the group/channel metadata obtained from Step 2. 
We use two QAnon-related keywords: \emph{qanon} and \emph{wwg1wga}.
The former refers to the conspiracy theory itself, while the latter is the QAnon movement's motto that refers to ``Where We Go One We Go All.''
We select these specific keywords mainly because they are prevalent and used extensively by members of the QAnon movement.
Overall, we find 204 Telegram groups/channels that include the above keywords on their group/channel metadata or any collected Twitter/Facebook posts.

\descr{4. Validating QAnon groups/channels.} 
Then, we validate that the selected groups/channels are related to QAnon and remove any groups/channels that are not directly related (i.e., mentioning QAnon only once because of mentions in the news). 
To do this, an author of this study, who has previous experience on the QAnon conspiracy theory, manually annotated the 204 groups/channels obtained from Step 3. 
Specifically, the annotator viewed each group/channel via Telegram's Web client and spent 5-10 minutes reading content shared in the group/channel and checking the group/channel metadata to decide on whether the group/channel is related to the QAnon conspiracy theory. 
Note that since many groups/channels are in languages other than English, the annotator used Google's translate functionality to translate content in English.
Overall, we annotate all 204 groups/channels and find 77 QAnon groups/channels.

\descr{5. Joining and collecting messages in QAnon groups/channels.} 
The next step in our data collection methodology is to join the QAnon groups/channels and collect all their messages.
We join all QAnon groups/channels, and then we use Telegram's API~\cite{Telegram-API} to collect all the messages shared within these groups.
Note that we only join and collect data from public groups/channels.
Initially, we collect 3.5M messages shared in 77 QAnon groups/channels from 92K senders between September 7, 2017, and March 9, 2021 (see Table~\ref{tab:dataset}).

\descr{6. Expanding QAnon groups/channels.} During our manual validation of the QAnon groups/channels, we observed many messages shared in QAnon groups/channels that are forwarded messages from other groups/channels.
Aiming to expand our set of QAnon groups/channels, we extract all groups/channels that forwarded messages in the 77 already discovered QAnon groups/channels and manually validate (see Step 4) the top 200 groups/channels in terms of the number of forwarded messages.
Note that we only validate the top 200, as manually checking and validating the groups/channels is time-consuming.
Using this approach, we discover an additional 84 QAnon groups/channels.
Then, we repeat Step 5 for the newly discovered groups and collect all their messages.
Overall, by combining the initial dataset and the one after expanding the QAnon groups/channels, we obtain a set of 4.5M messages from 92K senders in 161 QAnon groups/channels between September 7, 2017, and March 9, 2021 (see Table~\ref{tab:dataset}).

\descr{Baseline dataset.} We also collect a baseline dataset for comparing it with our QAnon dataset. 
To collect our baseline dataset, we follow Steps 1, 2, 3, and 5 as described above, with the only difference that we use a different set of keywords for selecting the groups/channels (note that we do not validate and manually check the groups/channels because they are not focusing on a specific topic).
Specifically, we use a set of keywords obtained from First Draft~\cite{firstdraft} that includes 133 keywords/phrases about important events that happened in 2020. 
For example, our keywords set includes words related to the 2020 US elections and the COVID-19 pandemic.
Overall, we join 869 channels and collect 9,358,946 messages shared between September 27, 2015 and  March 9, 2021 (see Table~\ref{tab:dataset}).

\descr{Ethical considerations.} 
Before collecting any data, we obtained approval from our institution's ethical review board. 
Also, we stress that: 
a) we work entirely with publicly available data; 
b) we do not make any attempt to de-anonymize users; and
c) we do not make any attempt to track users across platforms.
Overall, we follow standard ethical guidelines~\cite{rivers2014ethical} throughout our data collection and analysis.

\begin{figure*}[t!]
\centering
\subfigure[]{\includegraphics[width=0.3145\textwidth]{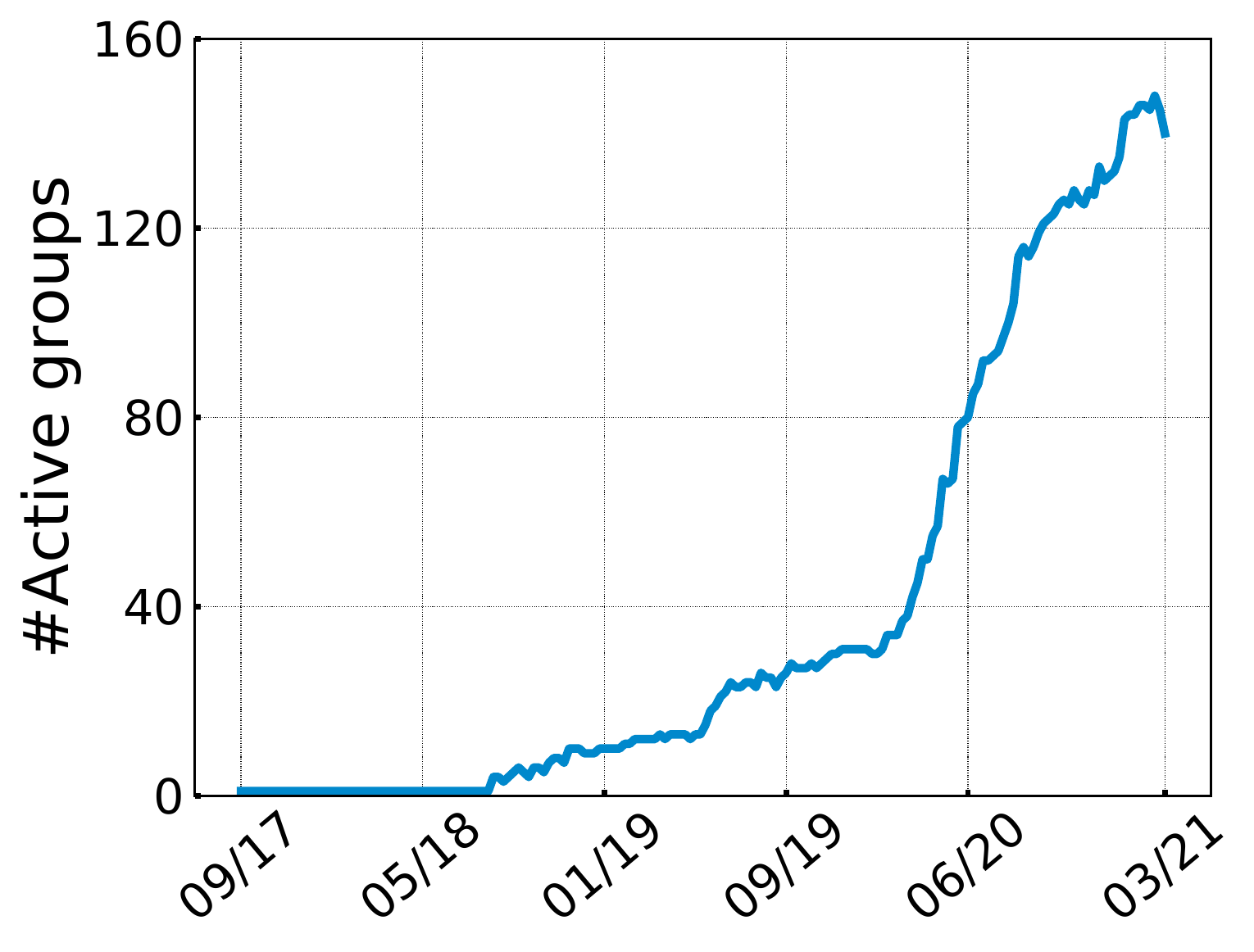}\label{fig:active_groups}}
\subfigure[]{\includegraphics[width=0.341\textwidth]{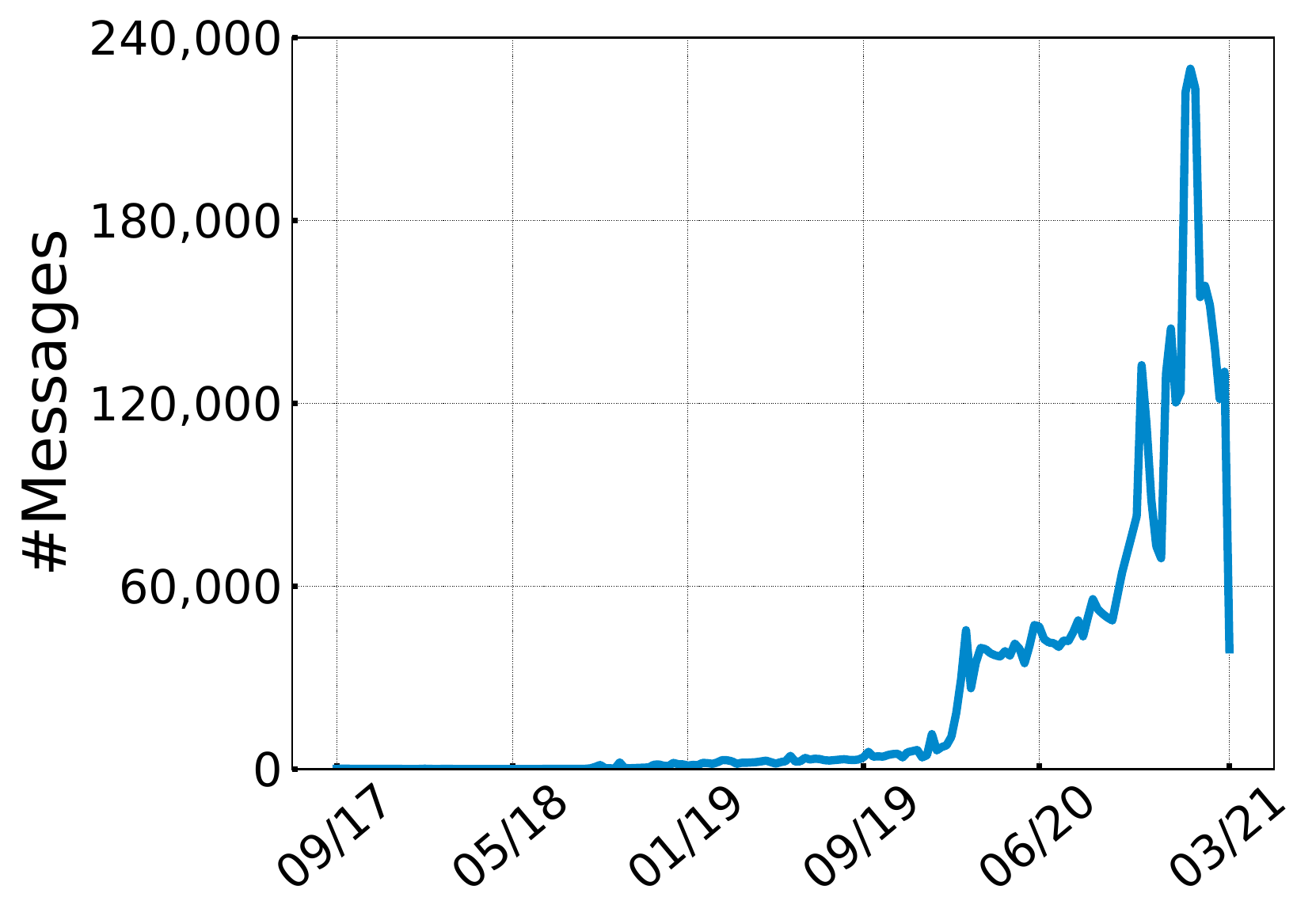}\label{fig:messages_per_week}}
\subfigure[]{\includegraphics[width=0.3345\textwidth]{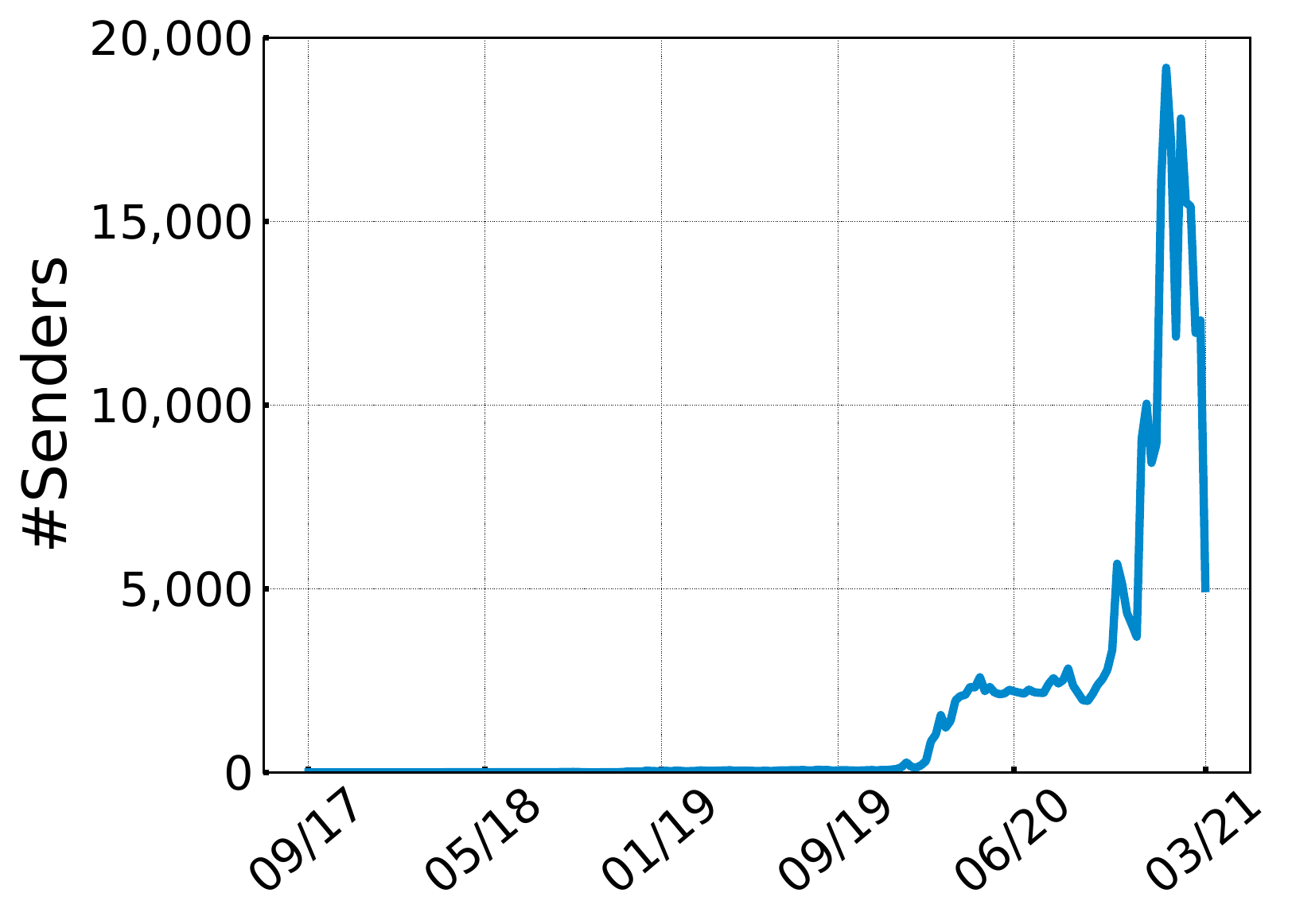}\label{fig:senders_per_week}}
\vspace{-4mm}
\caption{Activity within the QAnon groups/channels over time, in terms of active groups/channels, number of messages, and number of senders per week.}
\label{fig:temporal_activity}
\vspace{-4mm}
\end{figure*}

\section{Methods}\label{sec:methods}

In this section, we present our methods for analyzing multilingual content posted in QAnon-related groups/channels.

\descr{Topic Modeling.} One of the goals of this study is to analyze the QAnon-related discourse across languages.
To achieve this, we use a Bidirectional Encoder Representations from Transformers (BERT)-based topic modeling methodology by~\cite{grootendorst2020bertopic}.
Specifically, we use a pre-trained multilingual BERT model (stsb-xlm-r-multilingual) from~\cite{reimers2020making} to embed documents from multiple languages to the same high-dimensional vector space.
We select this specific model mainly because it supports 50+ languages and performs well in semantic similarity tasks.
Then, we use the Uniform Manifold Approximation and Projection (UMAP) approach proposed by~\cite{mcinnes2018umap} to reduce the dimensionality of the extracted embeddings.
This is an important step, as it allows us to reduce the dimensionality of the embeddings, hence increasing the performance and scalability of the next step (i.e., clustering).
Then, we group the reduced embeddings using the HDBSCAN algorithm~\cite{mcinnes2017hdbscan}.
Finally, using Term Frequency–Inverse Document Frequency (TF-IDF) on the clustered documents, we generate topic representations (i.e., a set of terms describing each topic). 

\descr{Toxicity Assessment.}
Also, we aim to quantify how extreme QAnon Telegram content is and whether there are changes in content's toxicity over time.
To assess this, we use Google's Perspective API~\cite{jigsaw2018perspective} to annotate each message in our dataset with a score that reflects on how rude or disrespectful a comment is.
Specifically, following previous work~\cite{ribeiro2020does}, we use the SEVERE\_TOXICITY model provided by the Perspective API and treat each message that has a score of 0.8 or more as toxic. 
We elect to use Perspective API for annotating content as toxic, mainly because it offers production-ready models that support multiple languages; as of May 2021, the Perspective API supports English, Spanish, French, German, Portuguese, Italian, and Russian.
Therefore, we can only assess the toxicity of messages posted in any of the seven languages above, which corresponds to 65\% of the messages in our dataset.
The rest of the messages do not include any text (20\% are sharing only audio, video, or images) or are in other languages (15\%) that the Perspective API does not support.
Note that the use of the Perspective API to assess the toxicity of content is likely to introduce some false positives or biases~\cite{davidson2019racial}. 
Previous work~\cite{gehman2020realtoxicityprompts}, has validated the performance of the Perspective API, however, it focuses mainly on the English annotations.
Therefore, although the models are in production, it is likely that there are biases in the Perspective API across languages.

\section{Results}\label{sec:results}

\subsection{Activity in QAnon Groups/Channels}

We start our analysis by looking into the general activity in QAnon groups/channels.
Fig.~\ref{fig:temporal_activity} shows the number of active groups, number of messages, and number of senders per week in our dataset.
Overall, we observe a substantial increase in the number of messages and senders after October 2020. 
This is likely because Facebook~\cite{facebook_suspensions} removed accounts and groups related to QAnon from their platforms during October 2020, hence users likely migrated to alternative platforms like Telegram.
Also, we observe a peak in activity during early 2021, which coincides with the attack in the US capitol by QAnon supporters. 
This initial analysis indicates that the QAnon conspiracy theory is growing rapidly on Telegram in terms of the number of groups/channels, the number of messages, and the number of users sharing messages.

\begin{figure}[tp]
	\centering
	\includegraphics[width=0.7\columnwidth]{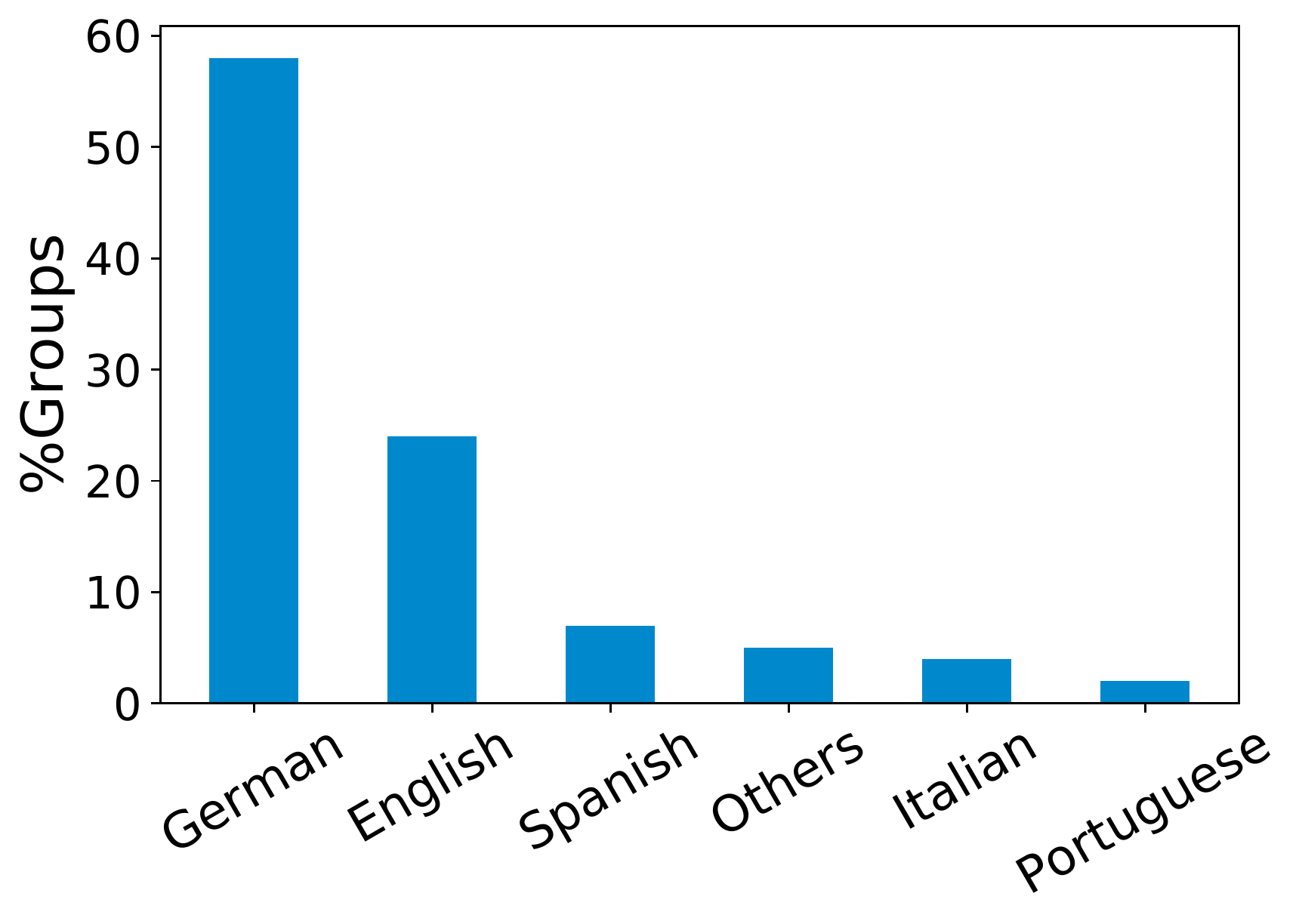}
	\caption{Percentage of groups/channels for each of the top languages (based on the language with the largest number of messages in each group/channel).} 
\label{fig:top-languages}	
\end{figure}

Next, we analyze the languages of QAnon content on Telegram.
Fig.~\ref{fig:top-languages} shows the percentage of groups/channels that have most of their messages in German, English, Spanish, Italian, and Portuguese (top five languages in our dataset).
Interestingly, we observe that the most popular language is German, with 58\% of the groups/channel sharing messages mainly in German.
The second most popular language is English with 24\% of the groups/channels, followed by Spanish with 7\%, Italian with 4\%, and Portuguese with 2\%.
Overall, these results highlight the multilingual aspect of the spread of conspiracy theories on platforms with a worldwide user base like Telegram.

\begin{figure}[t!]
\centering
\subfigure[]{\includegraphics[width=0.99\linewidth]{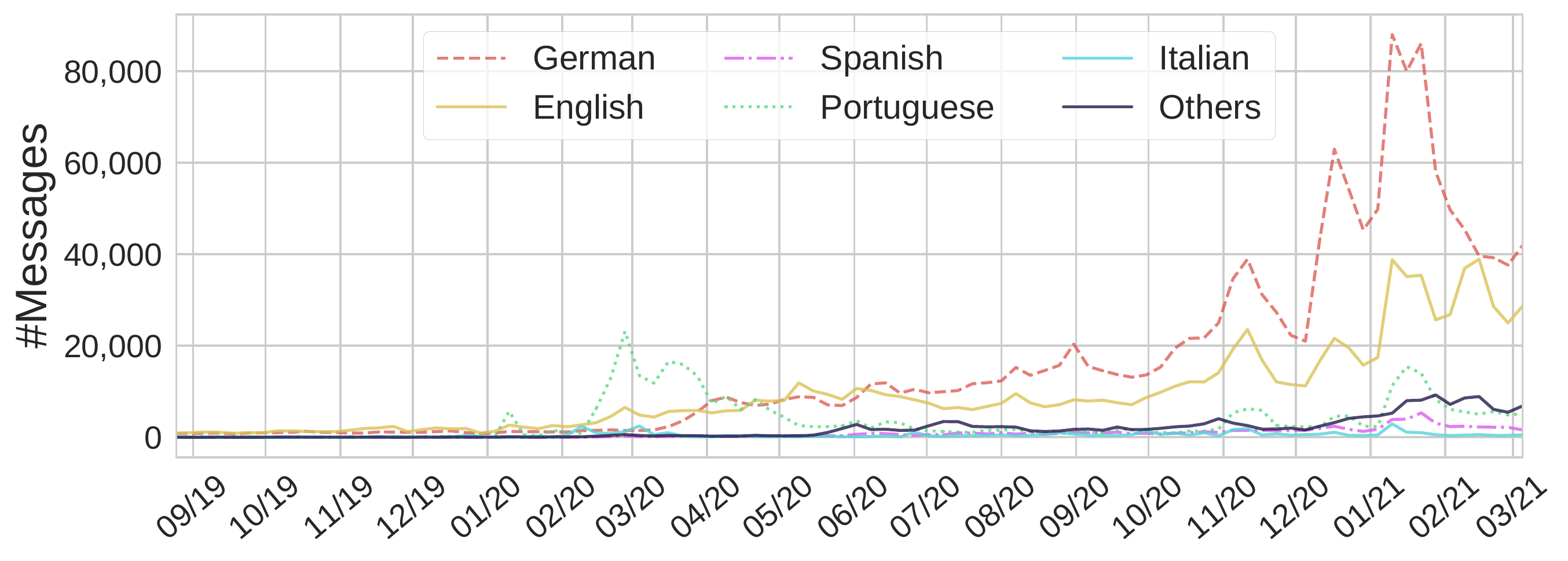}\label{fig:languages_over_time_abs}}
\subfigure[ ]{\includegraphics[width=0.99\linewidth]{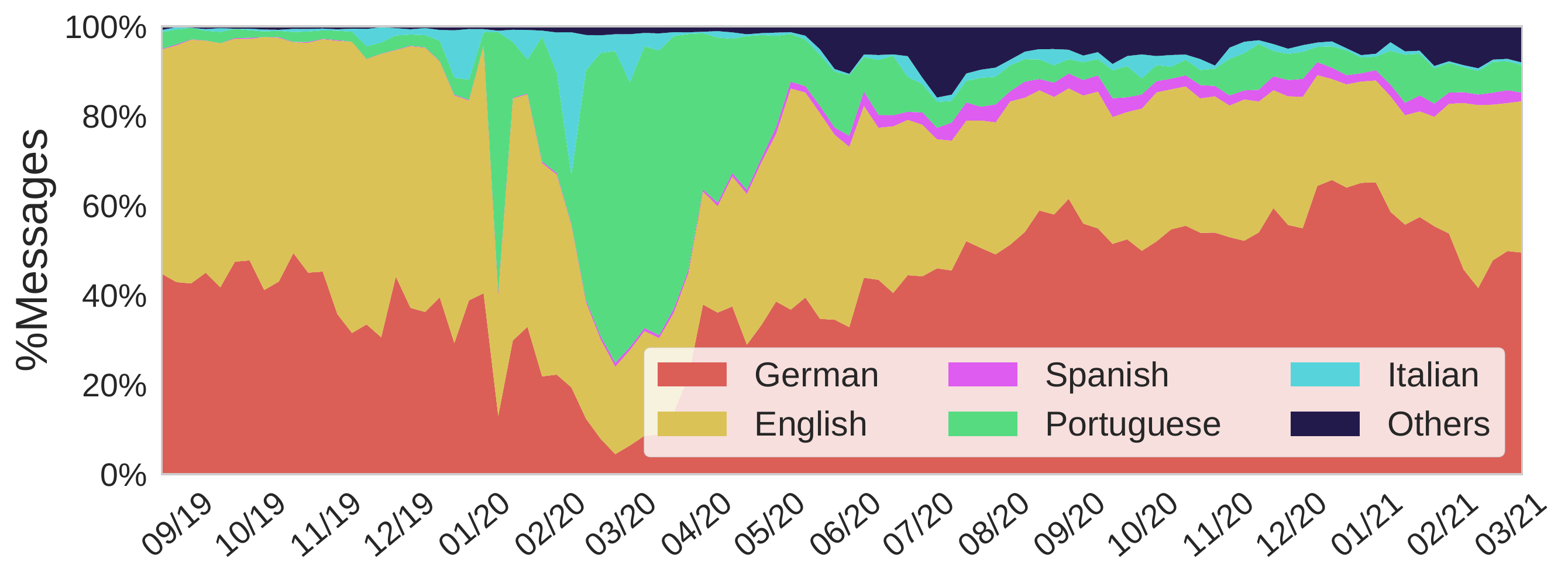}\label{fig:languages_over_time_stack}}
\vspace{-4mm}
\caption{Distribution of messages across languages on a weekly basis: (a) shows the number of messages for each language; and (b) the percentage of each language over the total number of messages.}
\label{fig:languages_over_time}
\vspace{-4mm}
\end{figure}

Next, we look into how the popularity of these languages changed over time to understand how the QAnon conspiracy theory became a global phenomenon on Telegram.
Fig.~\ref{fig:languages_over_time} shows the evolution of the languages over time on QAnon-related groups/channels (for readability purposes, we limit this analysis after September 2019).
Specifically, Fig.~\ref{fig:languages_over_time_abs} shows the absolute number of messages in each language per week, while Fig.~\ref{fig:languages_over_time_stack} shows the percentage of messages in each language weekly. 
We observe that English was the most popular language between September 2019 and December 2019, with over half of the QAnon-related messages posted in English, with German having a substantial percentage (around 40\% of all weekly QAnon related messages).
Then, between February 2020 and April 2020, we observe a substantial increase in the popularity of the Portuguese language, which became the most popular language for this period, overshadowing both English and German.
This period coincides with the beginning of the COVID-19 pandemic in Brazil when the virus was first confirmed to have spread to Brazil on February 2020~\cite{brazilian_1stcovid}.
Finally, after June 2020, we find that German is consistently the most popular language in our dataset, followed by English and Spanish/Portuguese having stable popularity.

\subsection{Toxicity in QAnon Groups/Channels}

Here, we investigate the toxicity of content shared within QAnon-related groups/channels on Telegram. 
The QAnon movement has links with events of real-world violence, hence it is important to analyze the toxicity of QAnon discussions on messaging platforms like Telegram.
We aim to uncover whether QAnon discussions on Telegram are more toxic than other discussions or QAnon discussions on other platforms and how toxicity changes over time in Telegram (i.e., are QAnon-related discussions becoming more toxic over time). 
To do this, as mentioned in Methods Section, we leverage Google's Perspective API that provides a signal (i.e., a score) of how toxic a message is.

\begin{figure}[t!]
\centering
\subfigure[]{\includegraphics[width=0.7\linewidth]{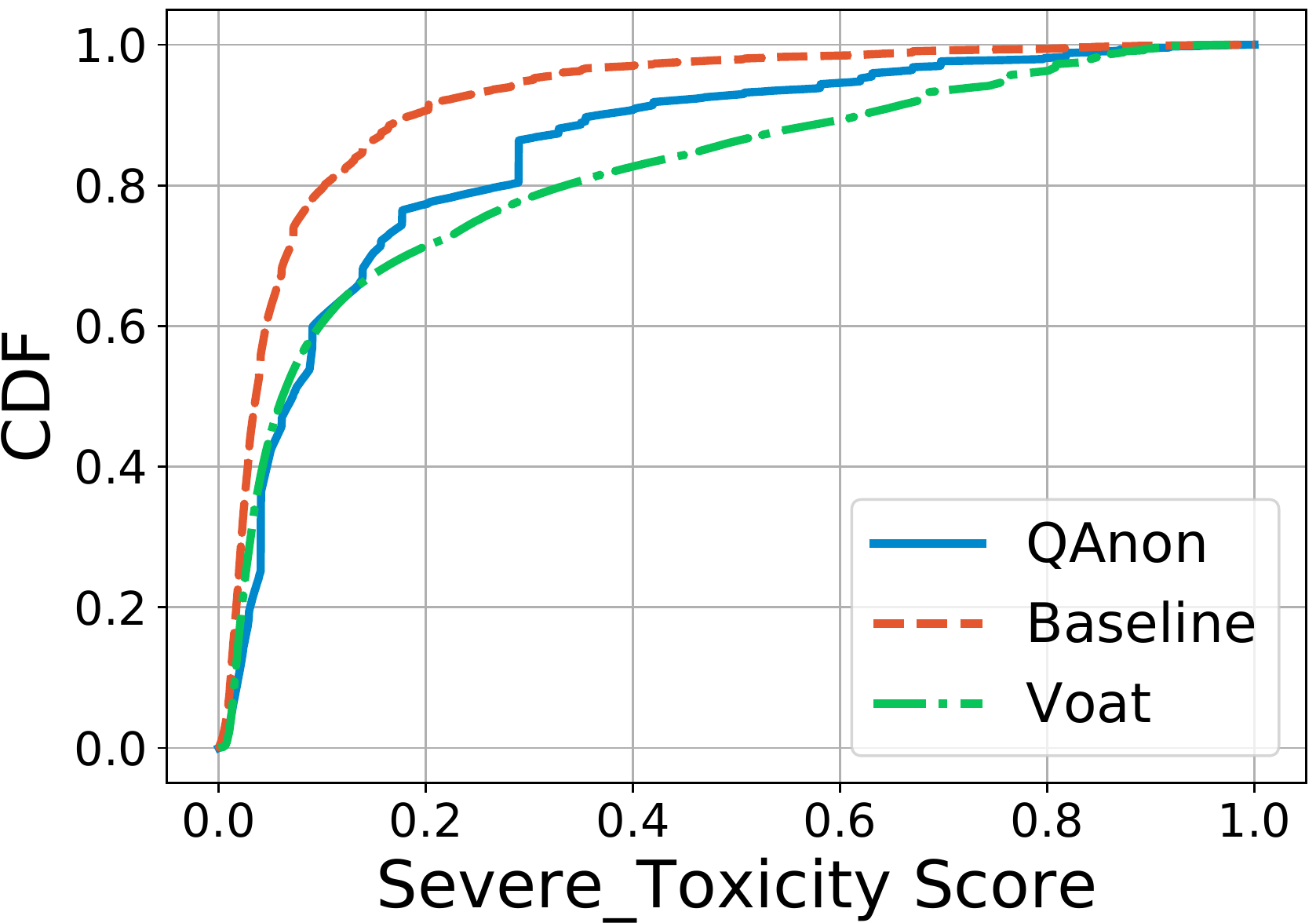}\label{fig:cdf_sever_toxicity}}
\subfigure[]{\includegraphics[width=0.7\columnwidth]{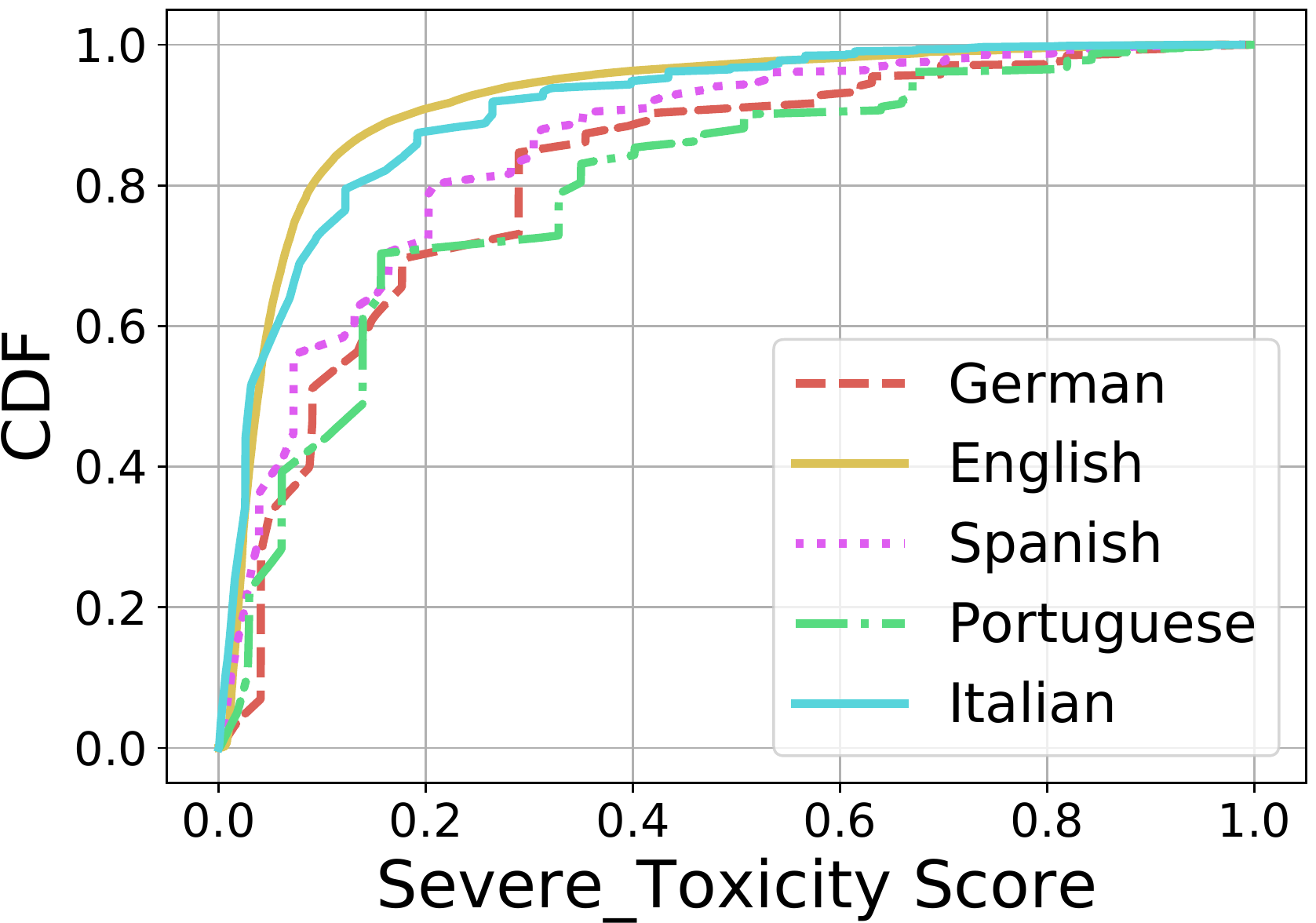}\label{fig:cdf_language_toxicity}}
\vspace{-4mm}
\caption{CDF of the toxicity scores obtained from Google's Perspective API. We compare toxicity scores in various datasets and across languages.}
\label{fig:toxicity}
\vspace{-4mm}
\end{figure}

First, we look into how toxic are QAnon discussions on Telegram by comparing it with our baseline dataset and the Voat dataset obtained from~\cite{papasavva2020qoincidence}.
Fig.~\ref{fig:cdf_sever_toxicity} shows the Cumulative Distribution Function (CDF) of the toxicity scores obtained from the SEVERE\_TOXICITY model in Perspective API.
Overall, we observe that QAnon discussions on Telegram tend to be more toxic compared to our baseline Telegram dataset (median score is {0.07} for QAnon discussions and {0.03} for baseline dataset). 
This finding is in contrast with the findings from~\cite{papasavva2020qoincidence} that found that QAnon discussions on Voat were less toxic compared to other popular content on Voat.
This difference in our findings and the findings from~\cite{papasavva2020qoincidence} is likely due to the fundamental differences that exist between the two platforms, while Voat is a fringe Web community mainly discussing conspiracy theories, Telegram is a more general-purpose and mainstream platform that includes less toxic non-QAnon-related discussions.
By comparing QAnon discussions across platforms (i.e., Telegram and Voat), we find comparable toxicity levels with QAnon discussions on Voat being very similar compared to Telegram (median toxicity score of {0.06} for Voat and {0.07} for Telegram).
To assess whether the distributions have statistically significant differences, we run a two-sample Kolmogorov-Smirnov test, finding that the distributions between Voat and Telegram are statistically significant ($p<0.01$).

\begin{figure}[t!]
	\centering
	\subfigure[]{\includegraphics[width=\columnwidth]{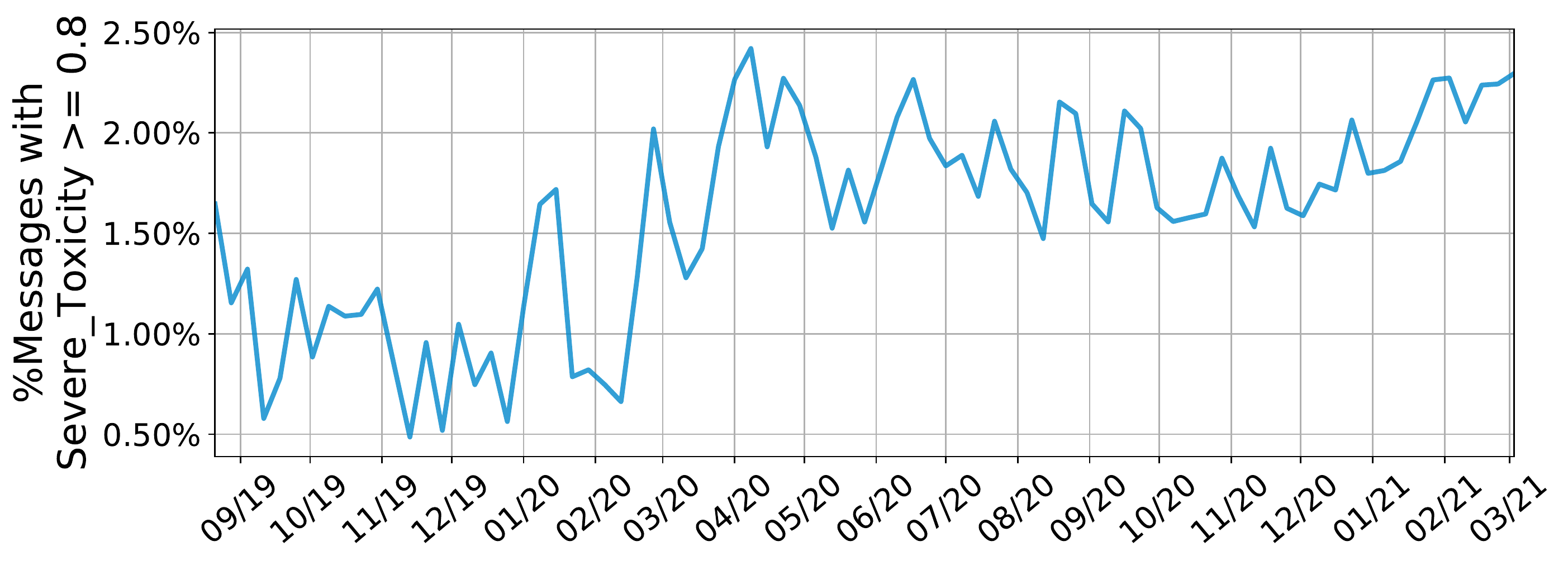}\label{fig:toxicity_over_time_all}}
	\subfigure[]{\includegraphics[width=\columnwidth]{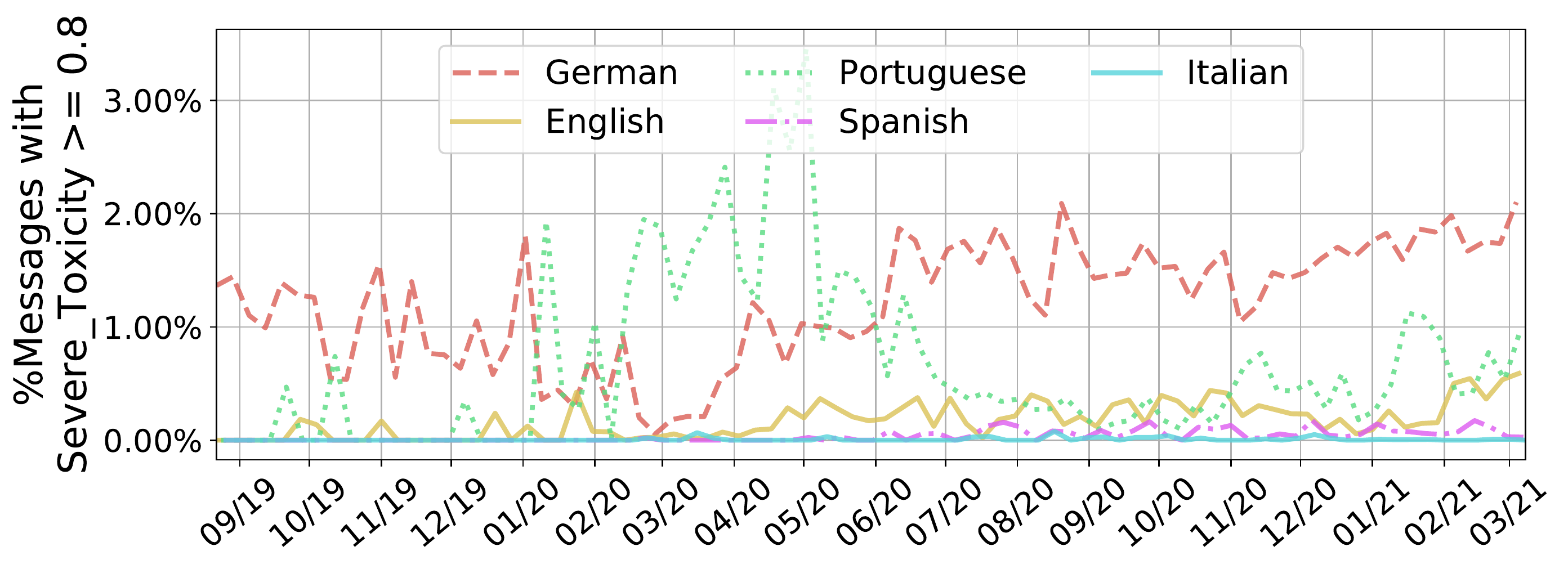}\label{fig:toxicity_over_time_language}}
	\vspace{-4mm}
	\caption{Percentage of messages that are toxic per week for: (a) all messages; and (b) for each of the top five languages in our dataset (i.e., percentage of messages that are toxic in specific language per week). } 
	\label{fig:toxicity-per-week}
	\vspace{-4mm}
\end{figure}

Previously, we observed that the QAnon movement has different popularity across languages over time, hence we aim to quantify whether toxicity in QAnon discussion varies across languages.
Fig.~\ref{fig:cdf_language_toxicity} shows the CDF of the SEVERE\_TOXICITY scores for all messages in the five most popular languages in our dataset.
Overall, we find that languages like German, Spanish, and Portuguese tend to be more toxic in QAnon discussions on Telegram compared to English and Italian.
Specifically, we find a median toxicity score of {0.09}, {0.07}, {0.14} for 
German, Spanish, and Portuguese, respectively, while for English and Italian, we find a median score of {0.03}.
Looking into the proportion of messages that are toxic (i.e., SEVERE\_TOXICITY score $>=0.8$, see Methods Section), we find {2.44\%}, {0.54\%}, {1.29\%}, {3.57\%}, {0.26\%} toxic posts, 
for German, English, Spanish, Portuguese, and Italian, respectively. 
Taken altogether, these results show substantial differences, in terms of toxicity, in QAnon discussions across various languages.
Also, by combining the findings from the previous section, our results indicate that the QAnon movement has become more popular in other languages over time (e.g., German), and at the same time, the discussion in these languages is more toxic.
These findings prompt the need to study such emerging issues by considering the multilingual aspect that exists in the spread of conspiracy theories that become a global phenomenon.

Finally, we look into how the toxicity of discussion changes over time.
Fig.~\ref{fig:toxicity_over_time_all} shows the weekly percentage of messages that are toxic for all the messages, while Fig.~\ref{fig:toxicity_over_time_language} shows the weekly percentage of toxic messages in a specific language for the five most popular languages in our dataset.
Our results show an overall increase in toxicity over time (see Fig.~\ref{fig:toxicity_over_time_all}).
Specifically, between September 2019 and December 2019, approximately 1\% of all the messages were toxic (per week). In contrast, by the end of our dataset, the same percentage rises to over 2\% of all the messages.
By looking into the toxic posts broken down to specific languages (see Fig.~\ref{fig:toxicity_over_time_language}), we observe an overall increase of toxicity by the end of our dataset, especially for German and English. 
For Portuguese, we observe a substantial increase in toxicity during early 2020, which coincides with the rapid increase in the overall number of messages posted in Portuguese (see Fig.~\ref{fig:languages_over_time_stack}).
By looking at the messages, we can observe those toxic messages are related to the COVID-19 pandemic in Brazil, including anti-vaccine conspiracies.
Also, we find politics-related messages that are attacking two Brazilian ex-ministers that left the government during this period.

\subsection{Popularity and Message Dissemination Across QAnon Groups/Channels}

This section discusses message forwarding across QAnon groups/channels and how popular is QAnon content on Telegram. 
Also, we explore how QAnon structures its network within Telegram groups/channels.

\begin{figure}[tp]
	\centering
	\includegraphics[width=0.7\columnwidth]{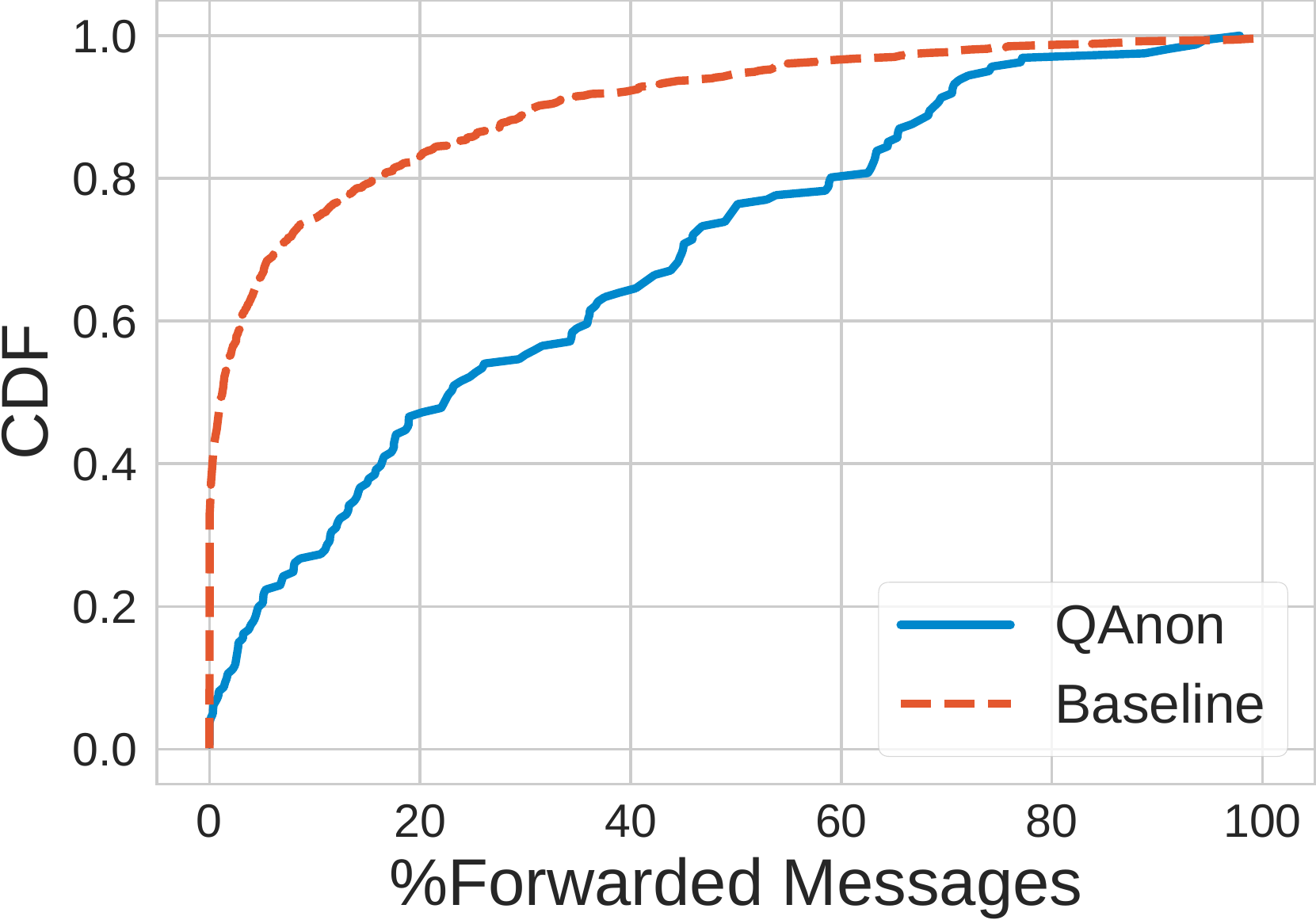}
	\caption{CDF of the percentage of messages that are forwarded per group/channel.} 
\label{fig:cdf_outter_source}	
\end{figure}

First, we evaluate the percentage of messages from each group/channel that are forwarded messages; i.e., messages that originate from another group/channel in Telegram. 
Fig.~\ref{fig:cdf_outter_source} shows the CDF of the percentage of forwarded messages per group/channel.
We observe that QAnon groups/channels tend to have more messages forwarded.
In 24\% of QAnon groups/channels, more than half of their messages are forwarded messages, while just 5\% of the groups/channels in the baseline have the same percentage of forwarded content.
Also, 48\% of the baseline groups/channels are rarely have forwarded messages (i.e., less than 1\% of all messages), while for QAnon groups/channels, only 8\% of them have similar behavior.

\begin{figure}[t!]
	\centering
	\includegraphics[width=\columnwidth]{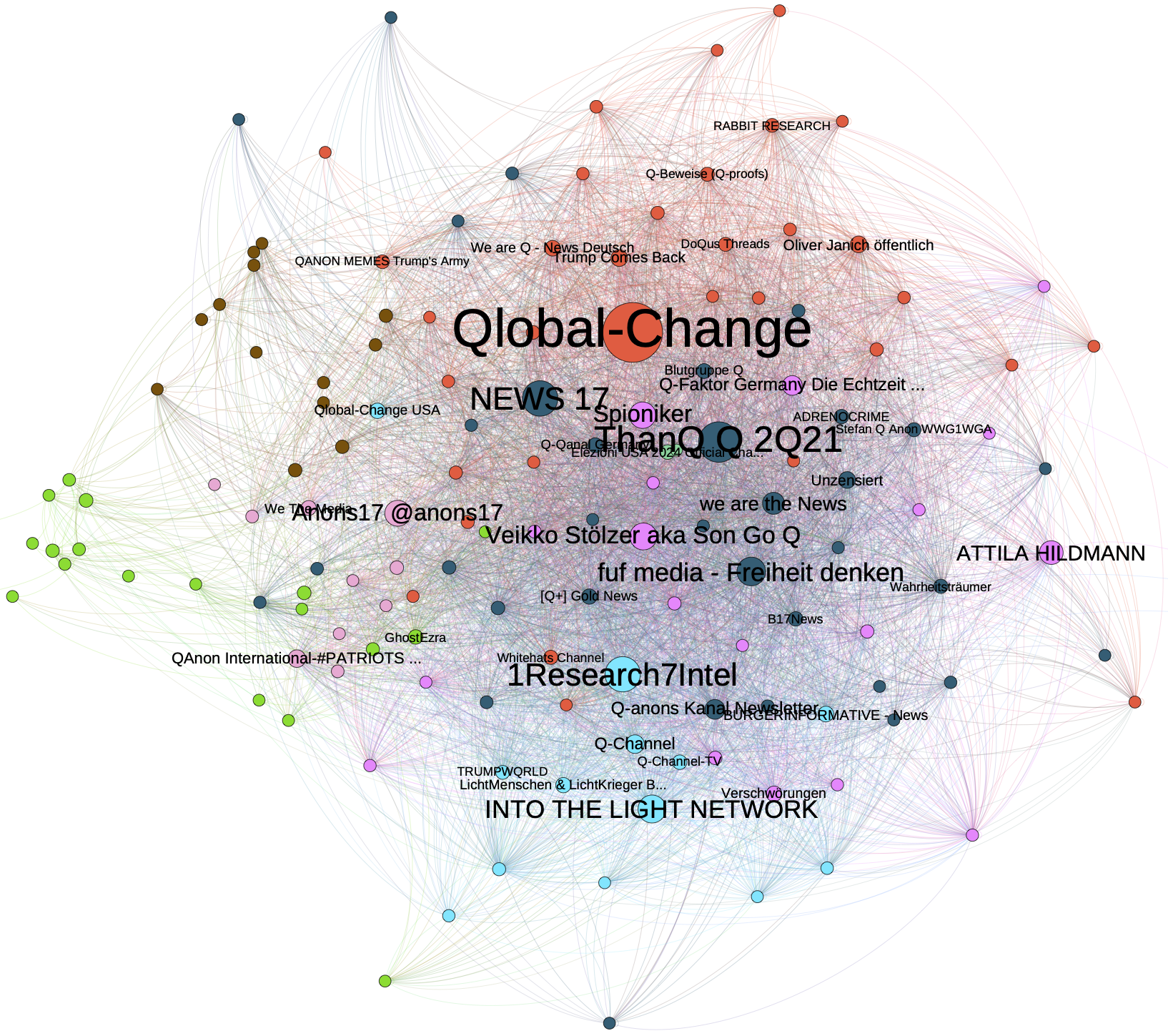}
	\vspace{-4mm}
	\caption{Forwarding graph representation of the QAnon groups/channels. Note that we omit the labels on small nodes for visualization purposes. } 
	\vspace{-4mm}
\label{fig:graph}	
\end{figure}

To better understand how messages are shared between QAnon groups/channels we visualize message forwarding graphs.
Fig.~\ref{fig:graph} shows a forwarding graph, where each node is a group/channel and an edge exists between them if they forwarded messages from one to another.
The size of the nodes is based on the number of messages that a node forwarded to other groups/channels and we use a visualization algorithm that takes into account the edge weights (in this case number of forwards), hence nodes with larger number of forwards are laid out closer in the graph space.
We also perform community detection using the Louvain method and color the nodes accordingly.
We observe that the graph consists of one huge connected component (90\% of the nodes), indicating that the QAnon community is a tightly-knit community that disseminates messages across multiple groups/channels.
Looking at the nodes with the biggest size, we observe groups/channels like ``Qlobal-Change'' that disseminates a large number of messages to other groups and in particular the nodes in the orange community (e.g., ``Trump Comes Back,'' ``QANON Memes Trump's Army,'' etc.)
Another community of interest is the pink one, which is seemingly an international set of groups with the groups ``Veikko Stoltzer aka the Son of Q'' and ``ATTILA HILDMANN'' forwarding a large number of messages to other international groups.
Also, the blue community at the bottom consists of groups/channel dedicated in QAnon research with groups/channels like ``1Research7Intel'' and ``INTO THE LIGHT NETWORK'' disseminating a lot of messages to other QAnon groups/channels.

\begin{figure}[t!]
	\centering
	\includegraphics[width=0.85\columnwidth]{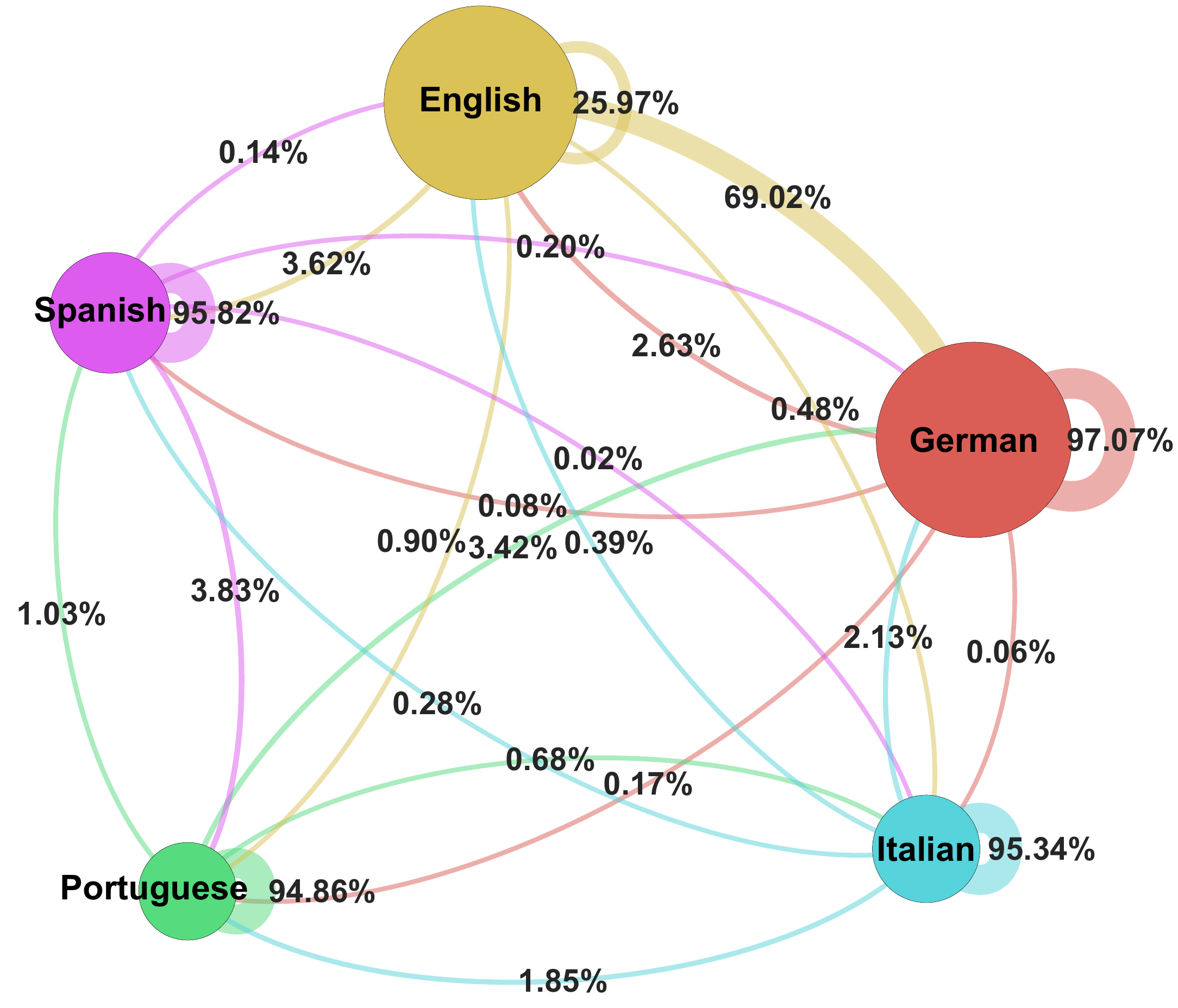}
	\caption{Forwarding graph with groups/channels aggregated based on their most popular language. The direction of the edge is based on the color of the source node and the percentage of messages is based on the overall number of forwarded messages from the source groups/channels.} 
	\label{fig:graph-langs}	
\end{figure}

We also visualize the forwarding graph by aggregating groups/channels based on their most popular language (Fig.~\ref{fig:graph-langs}), which allows us to quantify the interaction of QAnon groups based on their language.
We find that for all languages except English, the majority of forwarded messages (over 94\%) are within groups of the same language.
Interestingly for English groups/channels, 69\% of the forwarded messages are towards primarily German-speaking groups/channels.
Other non-negligible interactions between languages are Spanish to Portuguese (3.8\%), English to Spanish (3.6\%), and Portuguese to German (3.4\%).
Overall, these findings highlight interaction across languages within the QAnon community on Telegram, with the stronger connection being from English to German groups/channels.

\begin{figure}[t!]
\centering
\includegraphics[width=0.7\columnwidth]{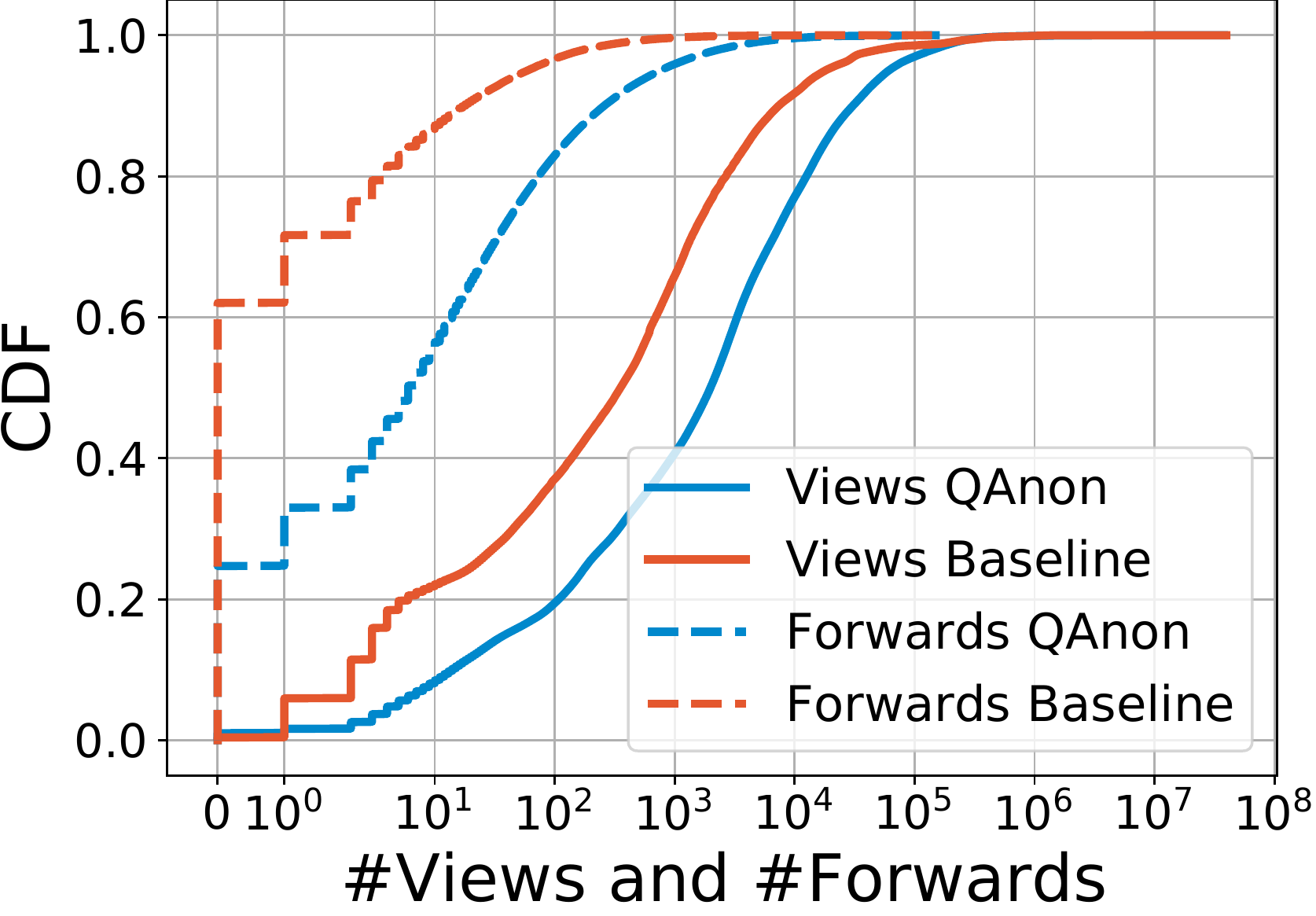}
\caption{CDF of the number of views and forwards (only for messages posted in channels).}
\label{fig:cdf_popularity}
\vspace{-4mm}
\end{figure}

\begin{figure}[tp]
	\centering
	\includegraphics[width=1.0\columnwidth]{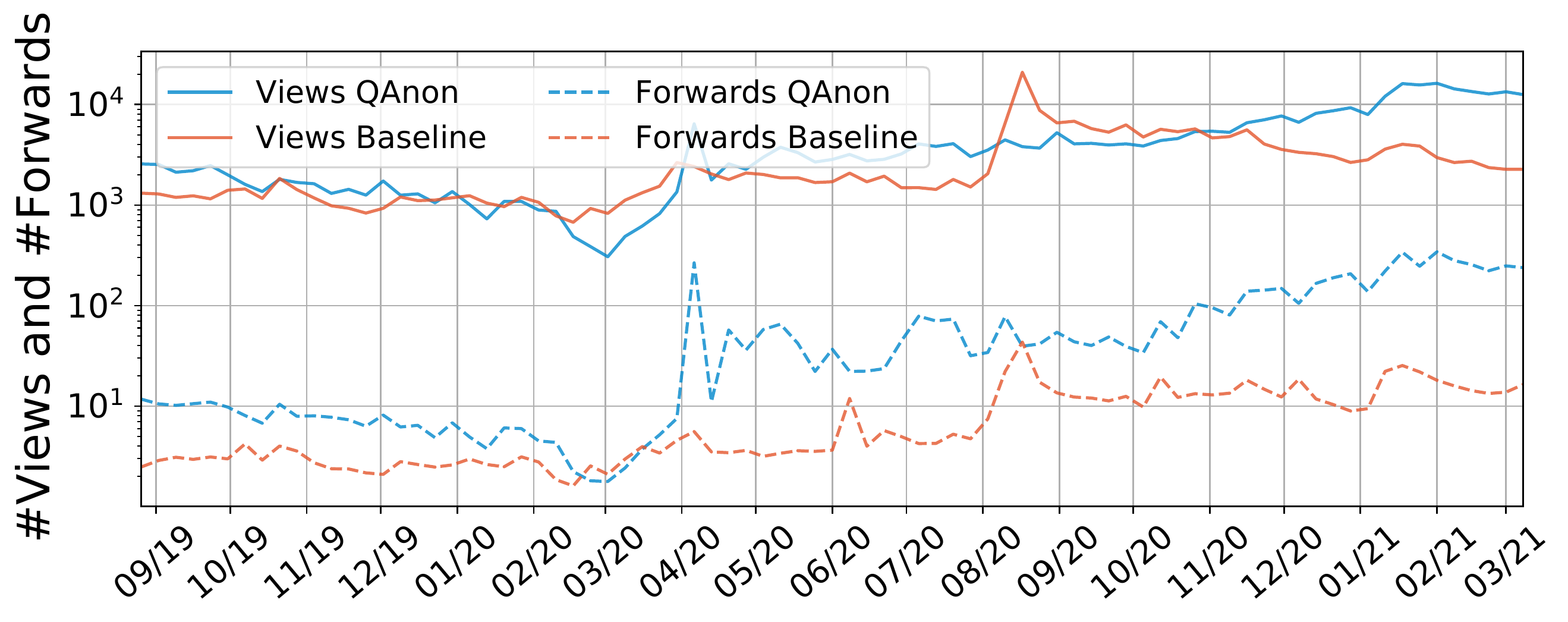}
	\caption{Mean number of views and forwards per week (only for messages posted in channels). } 
	\label{fig:views_fwd_day}	
	\vspace{-4mm}
\end{figure}

Next, we analyze the total number of views and forwards for the messages in our dataset, as they indicate how popular the QAnon content is across the \emph{entire} Telegram network (not limited to the set of groups/channels we collected).
Note that the Telegram API does not provide metrics about views and forwards for all messages; the metrics are available only for messages posted on channels, hence, here we report results for 52.31\% of the messages from our QAnon dataset and 38.61\% of the baseline messages.
Looking at Fig.~\ref{fig:cdf_popularity}, we observe that messages in QAnon groups/channels have a substantially larger number of views and forwards compared to the baseline dataset.
Specifically, we find a median of 1,986 views and 6 forwards for QAnon messages, while for the baseline dataset, we find a median of 236 views and 0 forwards.
These results suggest that QAnon content on Telegram is more popular than the baseline dataset as it has at least an order of magnitude more number of views and forwards across the entire Telegram network.
We also look into how the popularity of content changes over time; Fig.~\ref{fig:views_fwd_day} shows the mean number of views and forwards of the messages over time. 
We find a quick increase in both views and forwards for messages from the QAnon dataset in early April 2020 and still increasing in 2021, ending the period with a weekly mean of more than 100 forwards per message and more than 10,000 views per message (compared to the initial 10 forwards and 1,000 views per message in the 09/2019). 
This shows the increasing popularity of QAnon content over time and even surpassing the baseline dataset in one order of magnitude for both views and forwards at the beginning of 2021.

\subsection{QAnon Discourse Across Languages}

\begin{table*}[]
\centering
\resizebox{\textwidth}{!}{%
\begin{tabular}{@{}llrr@{}}
\toprule
\textbf{Topic}                                                & \textbf{Example Terms}                                                                                                                                                                                               & \textbf{\#Messages} & \textbf{\begin{tabular}[c]{@{}r@{}}\%Toxic \\ Messages\end{tabular}} \\ \midrule
COVID-19/Vaccines                                    & vacina,vaccine,impfen,impfungen,geimpft,vacinas,impfstoff,mandatory,vaccinations,nebenwirkungen & 63,508               & 1.98\%                                                             \\
Trump/US elections                                            & bless,gewinnt,gewonnen,win,won,supporters,trust,gewinnen,trump,siegen,presidents                                                                                                                                     & 44,567               & 1.11\%                                                             \\
Brazilian Politics                                            & brasileiro,paulo,rio,brasil,brasileira,jair,bolsonaro,governador,ministro,doria,                                                                              & 40,622               & 3.50\%                                                             \\
German Politics/Hitler & hitler,nazi,nazis,adolf,hitlers,alois,tochter,nazista,ss,nazareth,rothschild                                                                                                                                         & 27,428               & 4.99\%                                                             \\
Money/Crypto                                        & money,bitcoin,dinheiro,banken,geld,bank,schulden,gamestop,steuern,aktien                                                                   & 22,973               & 1.95\%                                                             \\
Deplatforming                                & instagram,account,gesperrt,suspended,deleted,accounts,banned,whatsapp,fb,facebook                                                                                               & 18,700               & 1.01\%                                                             \\
Religion                                                      & jesus,cristo,bibel,senhor,deus,palavra,gottes,christus,bible,religion                                                                                                                   & 18,173               & 1.52\%                                                             \\
QDrops/Promoting QAnon                                        & qs,qanons,qdrops,qmap,anon,clock,anons,posts,buchstabe,qmor,qq,missle,qposts,qr,publicaes                                                                                                                            & 18,047               & 0.51\%                                                             \\
Shootings/Guns                                                & suspect,shooting,shot,woman,charged,murder,police,cops,fatally,killed                                                                                                                  & 10,219               & 1.58\%                                                             \\
Italian Politics                                              & italien,italia,italy,conte,italienische,italienischen,italian,leonardo,salvini,italiano                                                                                                                        & 8,784                & 1.19\%                                                             \\
Jews                                                          & jews,jewish,juden,israel,judeus,rabino,judaism,zionisten,zionist,holocaust                                                                                                                          & 6,054                & 3.74\%                                                             \\
Satanists                                                     & satanisten,satan,satanic,satans,satanische,satanist,devil,satanismus,teufel,demnio,satanischen,diabo                                                                                                                 & 4,688                & 9.94\%                                                             \\
Plandemic                                                     & kits,positivo,plandemic,vacinada,test,hoax,cov,bito,protocola,covid1984lgen                                                                                       & 4,148                & 0.67\%                                                             \\
Pizzagate                                                     & pizza,pizzagate,anrufer,sir,gate,podesta,caller,comet,google,conspiracy                                                                                                            & 3,974                & 2.39\%                                                             \\
Obamagate                                                     & obamagate,michelle,obama,barack,2013,obamacare,hussein,prisoner,michae,geburtsurkunde                                                                                                        & 1,688                & 1.89\%                                                             \\ \bottomrule
\end{tabular}%
}
\vspace{-2mm}
\caption{Main topics extracted from our multilingual BERT topic modeling. We report the main theme of the topic, example terms that describe the topic, the number of messages that are mapped to each topic, and the percentage of those messages that are toxic.}
\label{tab:main-topics}
\vspace{-4mm}
\end{table*}

Thus far, we have looked into the evolution, toxicity, and popularity of QAnon content on Telegram, without analyzing the content itself. 
Therefore, here, we analyze the content of the messages shared within QAnon groups/channels with a focus on understanding the main topics of discussions and identifying differences across languages.  
To do this, as described in Methods Section, we employ a multilingual BERT-based topic modeling approach that enables us to generate topics across languages.

First, we preprocess all messages by removing emojis and URLs from the text and filtering out messages with an empty body (i.e., messages sharing only URLs, emojis, videos, or images).
After our preprocessing step, we end up with a set of 3M messages, which is the input for our topic modeling approach.
Then, using the topic modeling approach described in Methods Section, we extract a set of 2,023 topics. 
Note that 58\% of the messages are not mapped to any topic and are considered noise by the clustering algorithm.

\begin{figure}[t]
\centering
\subfigure[]{\includegraphics[width=\linewidth]{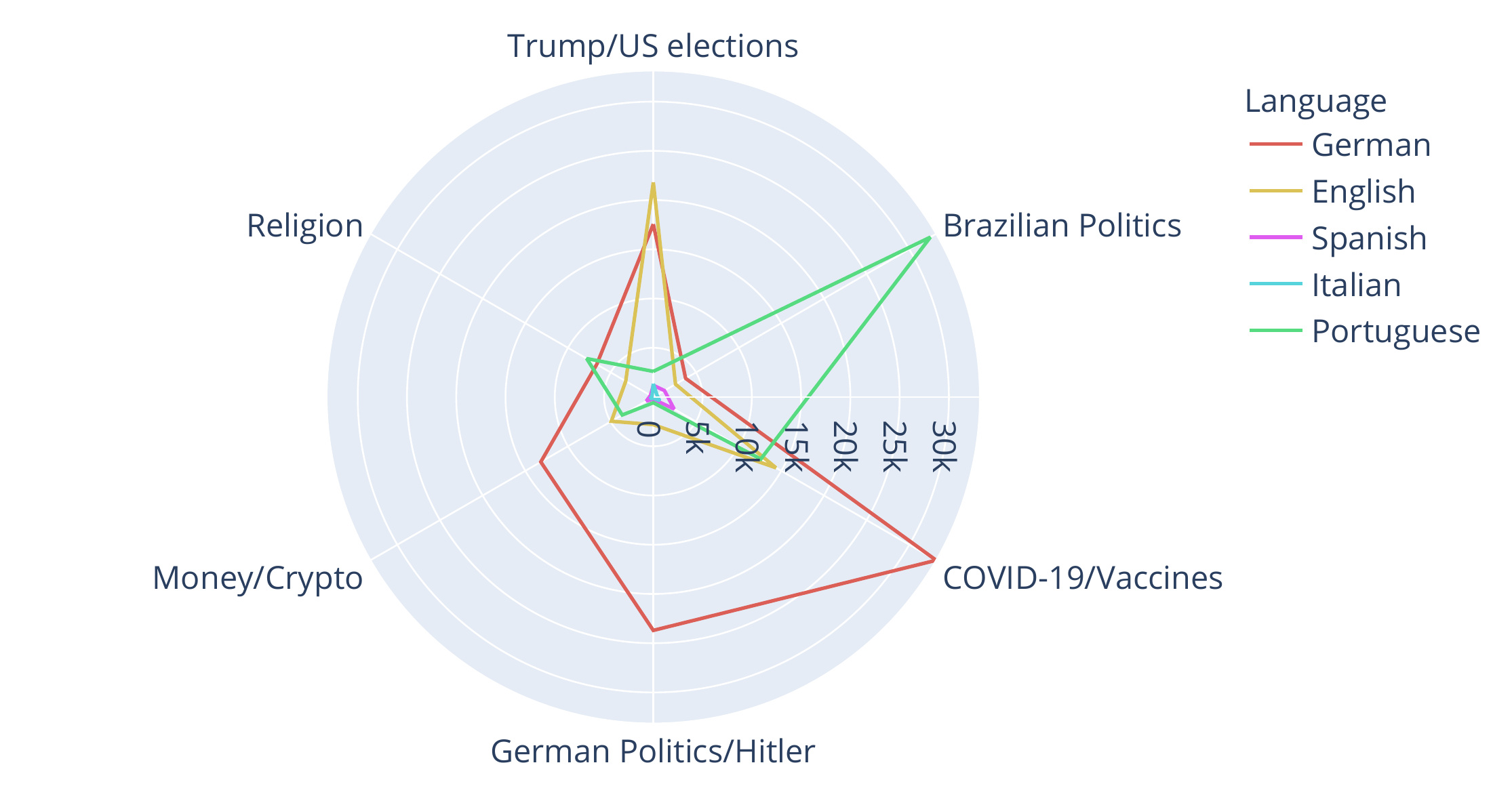}\label{fig:topics_languages}}
\subfigure[]{\includegraphics[width=\linewidth]{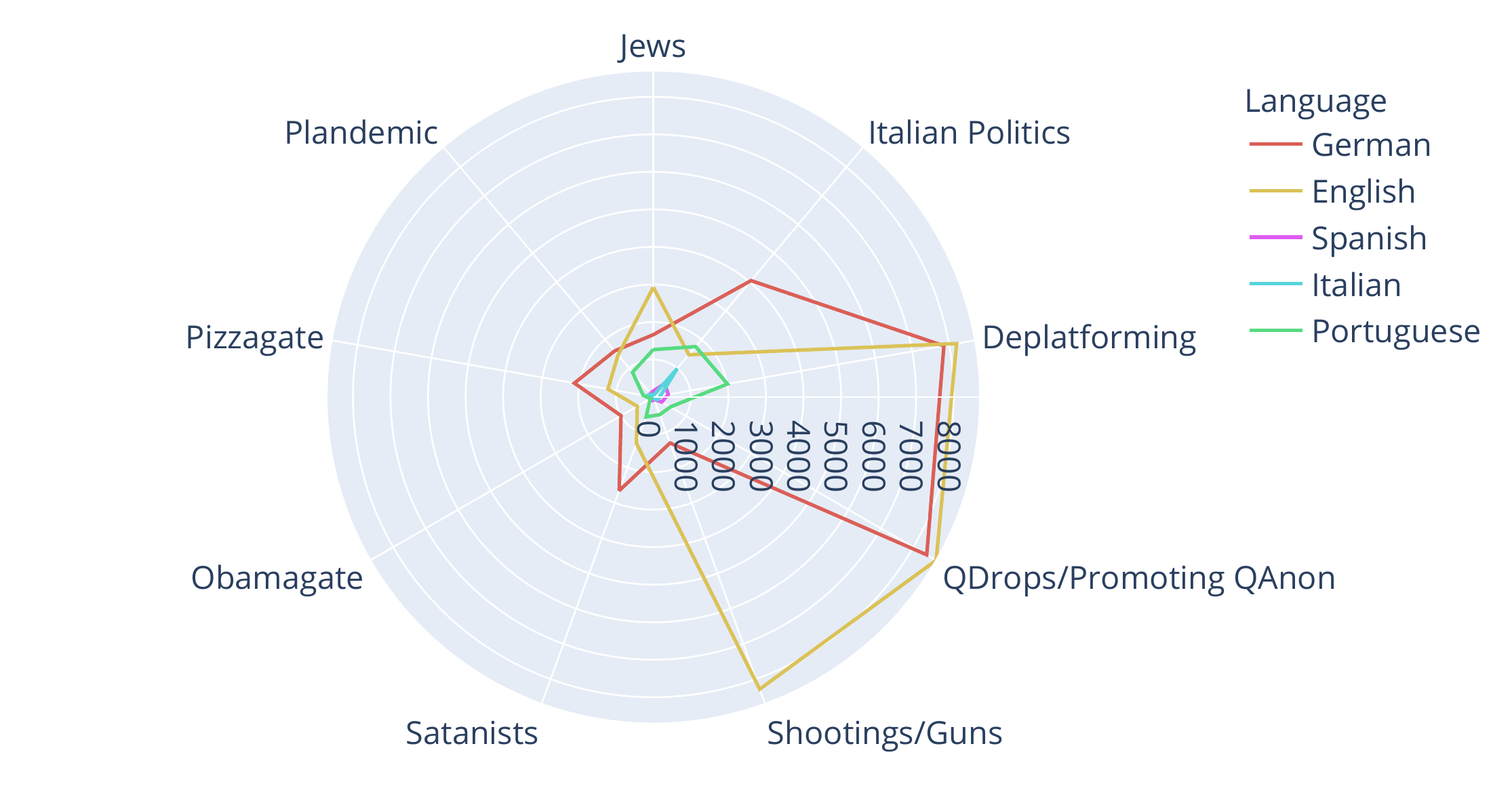}\label{fig:topics_languages2}}
\vspace{-4mm}
\caption{Distribution of topic-specific messages across languages.}
\label{fig:topics_languages_all}
\vspace{-4mm}
\end{figure}

Out of the 2K topics, we notice that many of them are related to the same high-level topic.
Therefore, we qualitatively analyze the top 100 topics in terms of frequency of messages to extract high-level topics.
Then, based on the keywords related to the high-level topics, we group together topics that are part of the same high-level topic.
Table~\ref{tab:main-topics} reports the extracted high-level topics from our analysis along with example terms (across languages), the number of messages that are mapped to each topic, and the percentage of those messages that are toxic, whereas Fig.~\ref{fig:topics_languages_all} shows the popularity of each topic across languages.

We make several interesting observations.
First, we find that the most popular high-level topic is related to the COVID-19 pandemic and discussions about the vaccines (63K messages).
Looking at the terms of this topic, we find the word ``nebenwirkungen'' (German) that translates to ``side effects'', hence Telegram users on QAnon groups were discussing possible side-effects from vaccines.
Overall, this topic is primarily popular among German groups and to a lesser extent to English and Portuguese ones (see Fig.~\ref{fig:topics_languages}).
Second, we find several topics about politics and real-world events, like the 2020 US elections, and German, Italian, and Portuguese politics.
This indicates that supporters of the QAnon conspiracy theory discuss political issues across multiple countries.
For German politics, we find several topics related to the Alt-Right and, in particular, discussions about Nazis and Hitler.
For all political topics, except Italian politics, their popularity across languages is based on the country of interest (see Fig.~\ref{fig:topics_languages_all}).

Other topics are discussions about money, cryptocurrency, and stocks (22K messages), deplatforming (18K messages), and religion (18k).
We also find specific topics that discuss Q drops and promoting QAnon content (18K), as well as discussions about fringe topics like various other conspiracy theories like the Plandemic (4K), Pizzagate (3.9K), and Obamagate (1.6K).
Also, similarly to~\cite{planck2020we}, we find discussions related to Satanism (4.6K messages), likely because QAnon supporters accuse the ``deep state'' as being Satanists.
Interestingly, this topic has the largest percentage of toxic messages, with 9.9\% of all messages mapped to this topic being toxic.

Additionally, we find a topic related to the Jewish community (6K messages), with 3.7\% of the messages in this topic being toxic.
Looking into the terms, we find the term ``holocaust,'' which indicates that QAnon supporters discuss the holocaust and the implications to the Jewish community.
The relatively high percentage of toxic posts on this topic likely indicates that QAnon supporters share antisemitic content.

\section{Conclusion}

In this work, we performed the first multilingual analysis of QAnon content on Telegram.
We joined 161 groups/channels on Telegram and collected a total of 4.5M messages shared over almost 4 years.
Using Perspective API and multilingual topic modeling, we shed light into how the QAnon conspiracy theory evolved and became a global phenomenon through Telegram.
Among other things, we found that the QAnon movement on Telegram is extremely popular in Europe, with German overshadowing English during 2020 and 2021.
Also, we find that QAnon discussions in other languages like German and Portuguese tend to be more toxic compared to English.
This indicates that not only the QAnon movement is becoming popular on other languages, it also becomes more toxic, similar to how viruses are mutating and become more lethal.
Additionally, we find that QAnon content is more popular on Telegram compared to a baseline of politics-oriented groups/channels, and that QAnon supporters discuss a lot of topics ranging from world politics, to COVID-19 vaccines, and various conspiracy theories. 
Taken altogether, our analysis highlights that the QAnon movement is adapting and becoming a political strategy that is embodied in far-right movements across the globe.
To conclude, our work highlights the need for the creation of organizations that aim to fact check and tackle the spread of QAnon-related false information across languages and countries (e.g., efforts similar to the \#CoronaVirusFacts Alliance focusing on the COVID-19 pandemic~\cite{covid_poynter}).

\small
\bibliography{references}

\end{document}